\documentclass[11pt,a4paper]{article}
\usepackage[T1]{fontenc}
\usepackage[utf8]{inputenc}
\usepackage{mathpazo}
\usepackage[scaled=0.95]{inconsolata}
\usepackage[margin=25mm]{geometry}
\usepackage{microtype}
\usepackage{graphicx}
\usepackage{booktabs,tabularx,array}
\usepackage{amsmath,amssymb}
\usepackage{enumitem}
\usepackage{xcolor}
\usepackage{caption}
\usepackage[authoryear,round]{natbib}
\usepackage{xurl}
\usepackage{hyperref}
\usepackage{placeins}
\usepackage{flafter}
\usepackage{xspace}
\usepackage{titlesec}
\usepackage{fancyhdr}

\definecolor{linkblue}{HTML}{245B87}
\newcommand{\reportauthors}{Alibaba Token Foundry, Alibaba Group}

\hypersetup{colorlinks=true,linkcolor=linkblue,citecolor=linkblue,urlcolor=linkblue,
  pdftitle={Qwen-Audio-Agent Technical Report},
  pdfauthor={\reportauthors}}
\newcommand{\sys}{Qwen-Audio-Agent\xspace}
\newcommand{\code}[1]{\texttt{#1}}
\newcolumntype{L}[1]{>{\raggedright\arraybackslash}p{#1}}
\titleformat{\section}{\large\bfseries\raggedright}{\thesection}{1em}{}
\titleformat{\subsection}{\normalsize\bfseries\raggedright}{\thesubsection}{1em}{}
\titlespacing*{\section}{0pt}{1.4em}{0.6em}
\titlespacing*{\subsection}{0pt}{1em}{0.35em}
\setlist{nosep,leftmargin=1.5em}
\begin{document}
\begin{center}
{\LARGE\bfseries Qwen-Audio-Agent Technical Report\par}
\vspace{0.8em}
{\normalsize\reportauthors\par}
\end{center}
\vspace{0.4em}
\begin{abstract}
We present \sys{}, a harness that combines full-duplex voice interaction with asynchronous task execution through a foreground--background architecture.
A Frontend Agent manages dialogue and selects between direct tool use and delegation, while a Backend Agent carries out delegated tasks in a separate context.
An Orchestration Runtime maintains task state, coordinates requests for user input and authorization, and schedules the return of results to the conversation.
The runtime separates speech interruption from task cancellation and execution completion from result delivery, allowing conversation to continue while delegated work proceeds.
Environmental events and persistent memory provide context within and across sessions.
Independent adapters support integration with different frontend models, backend agents, and clients.
We instantiate the architecture in desktop assistance, intelligent cockpits, and voice customer service.
On an in-house cockpit benchmark of 134 cases, mixed execution achieves a task success rate of 91.04\%, compared with 72.39\% and 80.60\% for the direct and all delegated configurations, respectively.
In a separate latency evaluation on matched successful turns, mixed execution reduces mean task execution latency by 26.73\% and 30.91\% relative to these baselines, respectively.
These results support the complementary use of direct tool calls for immediate operations and backend delegation for multi-step tasks.
\end{abstract}

\section{Introduction}
\label{sec:introduction}

Voice agents increasingly perform tasks that span several conversational turns.
A user may revise a request while a search is in progress, provide a missing constraint during a transaction, or ask about a document that is still being processed.
The agent must preserve task state across these exchanges while responding to new input.
For full-duplex voice agents, conversation, task execution, and result delivery must therefore be coordinated across different timescales.

Coordinating these activities requires separate control of conversation and task execution.
Interrupting a spoken response should not implicitly cancel ongoing work.
Similarly, a task may complete while the user is speaking, requiring its result to be retained until an appropriate point in the conversation.
We use \emph{harness} to denote the runtime infrastructure that connects models, tools, and execution environments and implements these interaction and task-management mechanisms.

Research on spoken turn-taking examines response timing and overlapping speech~\citep{skantze_turntaking}, while full-duplex models and benchmarks address simultaneous listening and speaking~\citep{moshi,fdb_original}.
Complementary work on language agents studies reasoning with environmental feedback~\citep{react}, multi-agent conversation~\citep{autogen}, and parallel tool execution~\citep{llmcompiler}.

Recent systems explore related forms of asynchronous voice interaction.
OpenAI's \code{gpt-realtime} supports asynchronous function calls~\citep{realtime_async}, while GPT-Live delegates reasoning and tool execution from a voice frontend to a backend~\citep{gpt_live_architecture,gpt_live_delegation}.
Thinking Machines Lab's Interaction Models pair an interaction model with an asynchronous background model~\citep{thinkingmachines2026interactionmodels}.
Gemini Live supports asynchronous function calls and, through its Extended Thinking model, background reasoning and tool use during spoken interaction~\citep{gemini_live_model,gemini_live_thinking}.
These developments motivate a common coordination layer for integrating different frontend models and backend execution systems.

\sys{} addresses this need through a common runtime for coordinating voice interaction and delegated tasks.
The key design choice is to manage tasks independently of conversational turns, allowing users to continue the dialogue while earlier work proceeds.
The runtime maintains task state and coordinates result delivery, while separate adapters support different frontend models, backend agents, and clients.
Environmental events and persistent memory maintain context within and across sessions.
The implementation is released as open source\footnote{Source code: \url{https://github.com/QwenAudio/qwen-audio-agent}.}.

Table~\ref{tab:system-comparison} compares the capabilities of \sys{} and related voice systems.

\begin{table}[!htbp]
\centering
\caption{Capability comparison of voice agent systems.}
\label{tab:system-comparison}
\small
\setlength{\tabcolsep}{5pt}
\renewcommand{\arraystretch}{1.16}
\begin{tabularx}{\linewidth}{@{}Xcccc@{}}
\toprule
\textbf{Capability} & \code{gpt-realtime} & GPT-Live & Gemini Live & \textbf{Qwen-Audio-Agent} \\
\midrule
Direct frontend tool use & $\checkmark$ & $\times$ & $\checkmark$ & $\checkmark$ \\
Dialogue during execution & $\checkmark$ & $\checkmark$ & $\checkmark$ & $\checkmark$ \\
Cross-provider frontend support & $\times$ & $\times$ & $\times$ & $\checkmark$ \\
Extensible backend agents & $\triangle$ & $\checkmark$ & $\triangle$ & $\checkmark$ \\
Backend task lifecycle management & $\triangle$ & $\triangle$ & $\triangle$ & $\checkmark$ \\
Extensible environmental events & $\triangle$ & $\triangle$ & $\triangle$ & $\checkmark$ \\
Persistent cross-session memory & $\triangle$ & $\triangle$ & $\triangle$ & $\checkmark$ \\
Open-source orchestration runtime & $\times$ & $\times$ & $\times$ & $\checkmark$ \\
\bottomrule
\end{tabularx}
\vspace{0.4em}
\begin{minipage}{\linewidth}
\footnotesize
$\checkmark$: provided; $\triangle$: requires application integration; $\times$: not exposed by the interface.
\end{minipage}
\end{table}


The report details task coordination, environmental events, and memory management, then examines their use in three application domains.
A cockpit benchmark compares direct, all delegated, and mixed configurations to assess the effects of execution routing on task success and latency.

\section{System Architecture}
\label{sec:architecture}

\subsection{Foreground--background architecture}
\label{sec:architecture-interfaces}

\begin{figure}[!htbp]
\centering
\includegraphics[width=\linewidth]{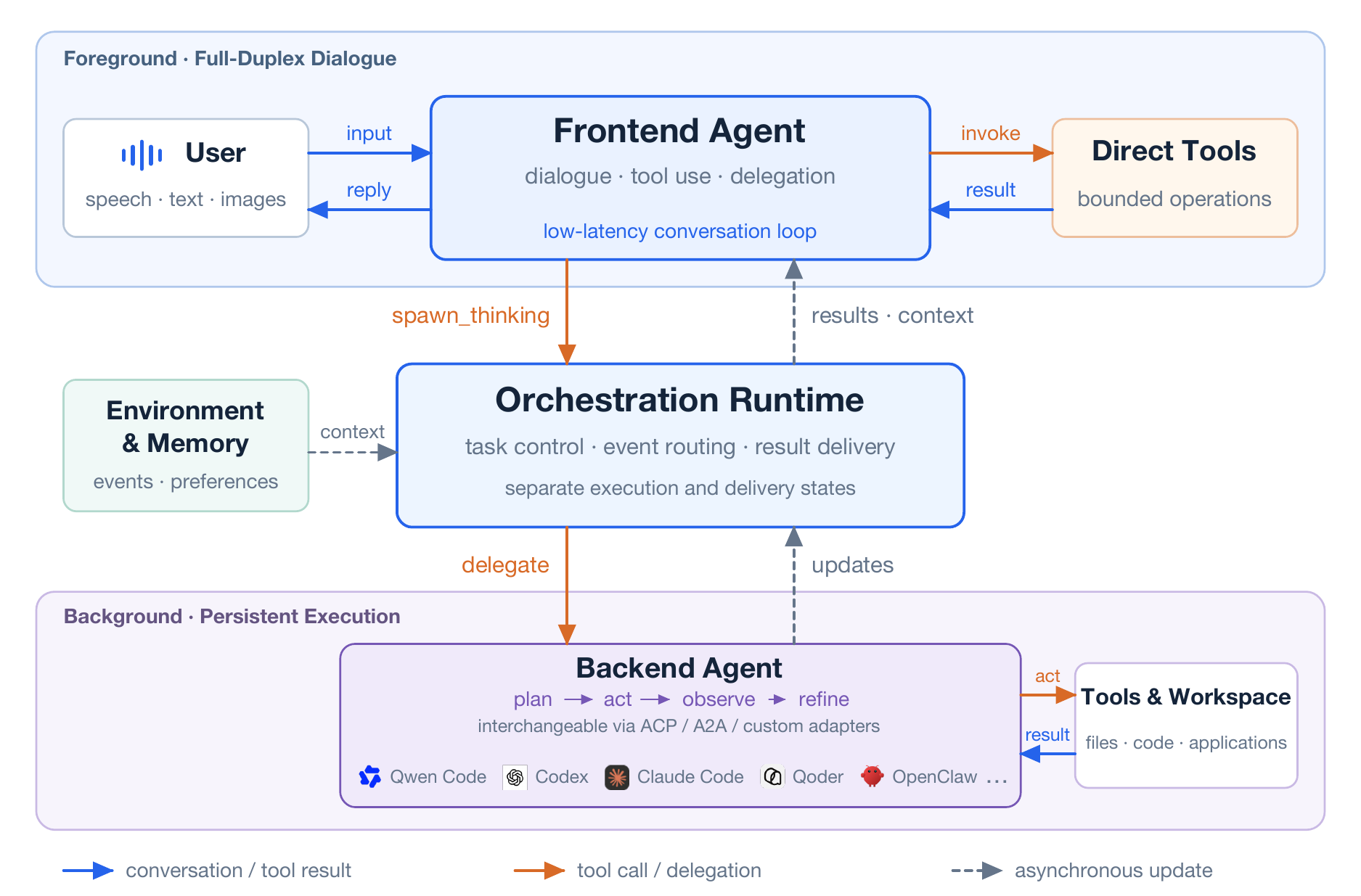}
\caption{Foreground--background architecture of \sys{}. The Orchestration Runtime coordinates full-duplex dialogue and asynchronous task execution. The backend examples illustrate interchangeable agent implementations.}
\label{fig:architecture}
\end{figure}

\sys{} separates ongoing conversation from delegated task execution (Figure~\ref{fig:architecture}).
The Frontend Agent interprets streaming input, maintains full-duplex dialogue, and chooses between direct tool calls and backend delegation.
The Backend Agent carries out delegated tasks using its own context, tools, and workspace.
The Orchestration Runtime tracks these tasks and routes progress updates, user input, and results between the two agents.
Clients capture input, play speech, display artifacts, and perform supported local actions.

The frontend uses direct tools for immediate operations and delegates tasks that require extended reasoning or a separate execution environment.
It selects an execution path according to the request and available capabilities.
Both paths remain available within a session, so immediate requests can be handled while delegated work continues.

Independent adapters allow frontend models and backend agents to be integrated through a common set of interfaces.
Frontend adapters translate streaming events and tool calls, while backend adapters expose task submission, progress, results, and supported controls.
Backend adapters use Agent Client Protocol (ACP)~\citep{acp}, Agent2Agent (A2A)~\citep{a2a}, or custom interfaces.
Each adapter declares the optional controls it supports, such as cancellation and requests for additional user input.

\subsection{Asynchronous task coordination}
\label{sec:coordination}

\begin{figure}[!t]
\centering
\includegraphics[width=\linewidth]{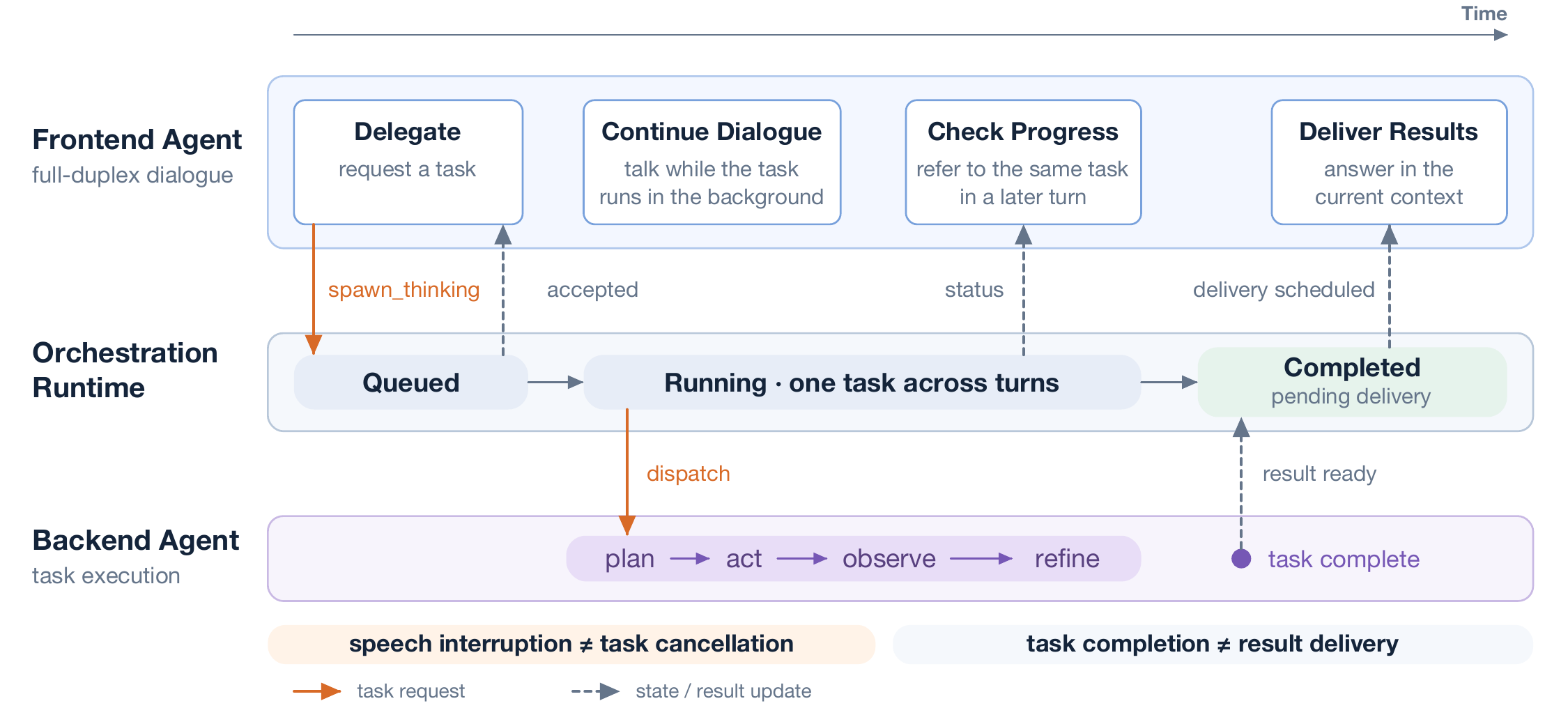}
\caption{Illustrative timeline of a delegated task. The Frontend Agent continues dialogue during execution, while the Orchestration Runtime tracks task state and schedules the completed result for delivery.}
\label{fig:coordination}
\end{figure}

Asynchronous delegation is exposed through the \code{spawn\_thinking} tool call.
The frontend submits a self-contained objective that includes the relevant constraints and input references.
This objective provides the context needed for execution, since frontend dialogue history and memory are not automatically transferred to the backend.
The runtime creates a task record and immediately acknowledges acceptance with a task identifier.
The frontend can then continue the conversation while the task is queued or running.
Figure~\ref{fig:coordination} illustrates how this task persists across subsequent conversational turns.

Each task record contains its objective, status, progress, outputs, and pending requests.
The common lifecycle distinguishes \code{queued} and \code{running} states from the terminal states \code{completed}, \code{failed}, and \code{cancelled}.
The frontend can query the same record in later turns.
When supported by the backend, requests for information or authorization remain linked to the originating task, allowing the runtime to route the user's response to the corresponding pending request.
Follow-up work can be submitted as a new task that refers to an earlier result.

Speech interruption and task cancellation are controlled independently.
Interrupting a response stops spoken output while preserving the task's execution state.
The runtime forwards task cancellation requests through the controls supported by the backend.
This distinction allows the conversation to be redirected without implicitly terminating unfinished work.

Result delivery is scheduled separately from execution completion.
The runtime retains completed results while the user is speaking or a frontend response is in progress.
Results that arrive within the same collection window are combined into one response request.
The frontend presents results in the current dialogue context, while artifacts remain linked to their tasks.
Separate execution and delivery records allow an unsuccessful delivery to be retried without repeating completed work.

\subsection{Environment perception and unified events}
\label{sec:environment}

Environmental context can change independently of spoken input.
Supported clients supply visual observations and application or device state, enabling the frontend to interpret subsequent requests against the current environment.
For example, when a user changes the destination through the cockpit interface, the new destination can be added to conversational context without triggering a spoken response.
Clients also report whether requested actions succeeded, providing execution feedback to the runtime.

IrisTK coordinates multimodal modules through standardized events~\citep{iristk}.
In \sys{}, application and device state changes use a unified event protocol, independent of the frontend provider.
Each event type defines a schema, retention rule, and response policy.
The runtime validates an event and either handles it directly, updates frontend context silently, schedules a response, or interrupts the current response and requests a new one.
Frontend adapters translate these actions into provider-specific messages.

\subsection{Memory and user context}
\label{sec:memory}

Prior work explores memory retrieval and reflection~\citep{generative_agents} and hierarchical memory management~\citep{memgpt}.
LongMemEval evaluates the retention and updating of information across sessions~\citep{longmemeval}.
In \sys{}, session context combines core interaction rules, an assistant profile, explicit user preferences, and long-term user memory.
The frontend uses a bounded memory snapshot to construct its context, while remote retrieval and consolidation can proceed asynchronously.
Task records and reference documents remain separate from personal memory.

The default memory implementation supports explicit edits during dialogue and extracts additional long-term instructions and stable facts at session boundaries.
An optional learning process adds inferred preferences only when supported by consistent evidence across sessions.
Current user requests take precedence over stored preferences, and explicit preferences take precedence over inferred preferences.
Revision checks prevent delayed background updates from overwriting newer corrections.
Preferences rejected by the user cannot be added again automatically.
These mechanisms adapt the context of later interactions without changing model parameters.

\section{Applications}
\label{sec:applications}

We instantiate \sys{} in desktop assistance, intelligent cockpits, and voice customer service. These applications place different demands on task coordination: iterative work on files, interaction with a changing environment, and transactions governed by policy and user approval.

\subsection{Desktop assistance}
\label{sec:applications-desktop}

Office and programming tasks often require inspecting a workspace, planning a sequence of actions, and revising the plan based on execution feedback~\citep{osworld,swe_agent}. The desktop implementation delegates these objectives to a backend agent with workspace tools, allowing document editing, code execution, and verification to proceed across conversational turns. During execution, the frontend remains available for dialogue and progress queries. Completed tasks return summaries and artifact references, while the full deliverables remain accessible through the workspace or client. Users can inspect and revise these artifacts through subsequent dialogue, linking spoken delegation to verifiable outputs.

\subsection{Intelligent cockpits}
\label{sec:applications-cockpit}

Cockpit interaction combines immediate control requests with longer tasks, while vehicle and application state can change through tool calls or direct interface actions. We address this requirement through a shared domain service: frontend tools, backend tools, and the graphical interface use the same business state and execution logic. Relevant state changes are returned as environmental events, keeping subsequent dialogue grounded in the current vehicle and application context.

The reference application assigns tools by domain: immediate controls and navigation to the frontend, and selected shopping and research tasks to the backend. Vehicle controls remain available while background work proceeds. In the benchmark, mixed execution instead selects adaptively between two paths with access to the same tools.

\subsection{Voice customer service}
\label{sec:applications-service}

Customer-service workflows require policy interpretation, access to business records, and user decisions at specific execution steps~\citep{tau_bench}. The retail and airline implementations combine policy retrieval with a shared business service whose tools and decision tables apply eligibility rules and calculate transaction amounts. Operations requiring approval follow a preview-and-commit sequence. The tool first prepares the proposed change, and the backend task pauses while the frontend obtains the user's decision. If the user approves, the task resumes and commits the operation. Linking this decision to the pending task preserves the execution context across conversational turns.

In addition to these three domains, \sys{} has also been deployed in digital-human and intelligent-hardware applications.

\section{Experimental Evaluation}
\label{sec:evaluation}

\subsection{Experimental setup}
\label{sec:evaluation-setup}

We compare four execution configurations on an in-house cockpit benchmark containing 134 cases: 86 short interactions and 48 multi-step tasks, spanning 159 user turns. All configurations use the same 41 business tools, initial states, and evaluation criteria. Qwen Audio 3.0 Realtime Plus serves as the Frontend Agent, with \code{qwen3.8-max} as the Backend Agent. The three audio configurations receive identical inputs and differ in execution routing: direct uses frontend tool calls, all delegated relies on the backend for all business-tool use, and mixed execution adaptively selects between the two paths. A text-only backend baseline receives the corresponding text and executes tools directly.

Evaluation combines adapted FDB V3 metrics~\citep{fdb_v3} with execution-based task success, consistent with the emphasis on task outcomes in interactive agent benchmarks~\citep{tau_bench,tau_voice}.

\subsection{Task completion}
\label{sec:evaluation-capability}

\begin{table}[!htbp]
\centering\footnotesize
\caption{Results on the in-house cockpit benchmark. The upper panel reports capability on 134 cockpit cases: F1 uses a 0--1 scale, and the remaining metrics are percentages. The lower panel reports mean latency on a separate set of turns completed successfully by all configurations. Bold denotes the best value in each column.}
\label{tab:execution-results}
\label{tab:capability}
\label{tab:latency}
\setlength{\tabcolsep}{5pt}
\renewcommand{\arraystretch}{1.06}
\begin{tabular}{@{}lcccc@{}}
\toprule
\raisebox{-1.8ex}[0pt][0pt]{Configuration} & \multicolumn{4}{c}{Capability evaluation} \\
\cmidrule(l){2-5}
& Tool F1 & Argument accuracy & Task success & Reply quality \\
\midrule
Audio frontend, direct tools & 0.947 & 93.6 & 72.39 & 95.5 \\
Audio frontend, all delegated & 0.908 & 89.8 & 80.60 & 89.6 \\
Text backend, direct tools & \textbf{0.963} & 94.1 & 90.30 & \textbf{100.0} \\
Audio frontend, mixed execution & 0.962 & \textbf{95.2} & \textbf{91.04} & 97.8 \\
\bottomrule
\end{tabular}

\vspace{0.6em}
\begin{tabular}{@{}lccc@{}}
\toprule
\raisebox{-1.8ex}[0pt][0pt]{Configuration} & \multicolumn{3}{c}{Mean task execution latency (s) $\downarrow$} \\
\cmidrule(l){2-4}
& Short tasks (74 turns) & Multi-step tasks (6 turns) & Overall (80 turns) \\
\midrule
Audio frontend, direct tools & \textbf{1.488} & 67.705 & 6.455 \\
Audio frontend, all delegated & 4.266 & 38.653 & 6.845 \\
Text backend, direct tools & 3.149 & \textbf{35.449} & 5.572 \\
Audio frontend, mixed execution & 1.591 & 43.430 & \textbf{4.729} \\
\bottomrule
\end{tabular}
\end{table}

Table~\ref{tab:execution-results} shows that mixed execution achieves the highest task success rate at 91.04\%. Direct frontend execution succeeds on 83 of 86 short cases but only 14 of 48 multi-step cases. The all delegated configuration improves multi-step completion to 37 cases, while short-case completion falls to 71. Mixed execution completes 84 short and 38 multi-step cases, combining strong performance on immediate operations with the backend's ability to execute longer action sequences.

\subsection{Task execution latency}
\label{sec:evaluation-latency}

Latency is measured separately on 80 user turns from 71 cases, with all four configurations completing each turn successfully. Measurement begins at the end of the spoken request, or at text submission, and ends at task completion; subsequent reply generation and playback are excluded.

Mixed execution achieves the lowest overall mean latency at 4.729\,s, a reduction of 26.73\% relative to direct frontend execution and 30.91\% relative to the all delegated configuration (Table~\ref{tab:execution-results}). On these turns, mixed execution routes all short tasks to the frontend and all multi-step tasks to the backend. It retains low latency on short requests while reducing multi-step latency relative to direct frontend execution. The all delegated configuration and the text-only backend baseline are faster on multi-step turns.

\FloatBarrier
\section{Conclusion}
\label{sec:conclusion}

\sys{} extends voice interaction to tasks whose execution continues across conversational turns.
Explicit task management and scheduled result delivery allow ongoing dialogue and background work to coexist within the same interaction.
Independent integration interfaces make this architecture applicable to different models, backend agents, and application domains.
The cockpit evaluation supports retaining both execution paths: mixed execution achieves higher task success and, on matched successful turns, lower overall mean execution latency than the direct and all delegated configurations.
Future work will examine conversational behavior during task execution, recovery over longer sessions, and personalization across a broader range of models, agents, and devices.

\FloatBarrier
\section{Contributions and Acknowledgments}
\label{sec:contributions}

All authors of \sys{} are listed in alphabetical order by last name.

\textbf{Authors:} Chong~Deng, Yunjie~Ji, Yuxiang~Kong, Xiangang~Li, Xu~Li, Binbin~Zhang, Haina~Zhu, and Jianheng~Zhuo.

\clearpage
\bibliographystyle{plainnat}
\bibliography{references}
\end{document}